\documentclass[prl,twocolumn,amsmath,amssymb,showpacs,aps]{revtex4-1}
\usepackage{graphicx}

\newcommand{\ket}[1]{ {\vert#1\rangle}}
\newcommand{\bra}[1]{ {\langle#1\vert}}
\newcommand{\braket}[2]{ {\langle#1\vert#2\rangle}}

\newcommand{\GS}{\Omega}
\begin{document}

\title{Universality in entanglement of quasiparticle excitations}
\author{Iztok Pi\v{z}orn}
\affiliation{
Vienna Center for Quantum Science and Technology, 
Faculty of Physics, University of Vienna, Boltzmanngasse 5, A-1090 Wien (Austria)}
\date{\today}

\begin{abstract}
We show that the entanglement entropy of single quasiparticle excitations of one dimensional systems exceeds the ground state entanglement entropy for $\log(2)$, if the correlation length of the system is finite. For quadratic fermion systems we show that the excess of entanglement is related to the number of quasiparticles in the excited state. This observation is confirmed by numerical examples of one dimensional quantum many-body systems, including nonintegrable. Tensor network methods to describe quasiparticle excitations are discussed.
\end{abstract}

\maketitle

The entanglement is of central interest in quantum information theory, quantum computation and plays an important role in quantum many-body physics.  On one hand,
it is a valuable resource in quantum computation and the reason why quantum algorithms can be  more powerful than classical ones \cite{nielsenbook}. On the other hand, it makes classical simulation of quantum many-body systems more difficult in terms of tensor network methods relying on a low degree of entanglement, such as the density matrix renormalization group \cite{white} or matrix product states \cite{vidalmps,frankmps} where the number of parameters scales roughly exponentially with the entanglement entropy \cite{schuch}.
The entanglement of quantum many-body systems at zero temperature has been extensively studied and, in particular for one dimensional lattice systems, it was shown that the ground state is much less entangled than generic quantum states, which makes a classical simulation tractable \cite{hastings}, and on the other hand provides a signature of quantum critical behavior \cite{osterloh} where the entanglement entropy violates the area law \cite{eisertreview}, connected to the central charge of the corresponding quantum field theory \cite{latorre,calabresecardy,korepin}. 
Much less, however, is known about the entanglement of excited states. It was shown by Alba, Fagotti and Calabrese \cite{alba} that entanglement entropies of excited states of integrable systems appear in a band-like structure and their ratios with the ground state entanglement entropy scale either extensively or logarithmically with the block size. 
For a critical Ising model, an exact scaling relation was obtained by Berganza, Alcaraz, and Sierra \cite{berganza} using a field theoretical approach. 
Still, no indication has been given that the banded structure of the \emph{excess of entanglement}, as the difference from the ground state entanglement entropy is called in \cite{berganza}, comes from the quasiparticle picture at low temperatures where a quasi particle causes an excess of $\log(2)$ to the entanglement entropy as compared to the ground state. This is what we show in the present Letter, assuming a translation invariance, a finite correlation length and a validity of a quasiparticle description at low energies. For quadratic models we also show that in the thermodynamic limit, the entanglement entropy of $k$-quasiparticle excitation exceeds the entanglement entropy of the ground state for $k \log(2)$.
The observation is supported by numerical simulations of spin chains: the Heisenberg XY model, the Ising model in a transverse field and the Ising model in a mixed field, the last being a nonintegrable model \cite{mussardo}.
Finally, we show that the excess of entanglement for the Bloch-like Ansatz for excited states \cite{jutho} is exactly equal to $\log(2)$ and is as such well suited to describe single particle excitations.

Let us consider a one-dimensional quantum lattice system of length $n$ described by a Hamiltonian $\hat{H}$ which is invariant to spatial reflections, $[\hat{H}, \hat{R}] = 0$, and conserves a parity of the system, $[\hat{H}, \hat{P}] = 0$. 
%For simplicity we assume that $\hat{H}$ is real. 
In a symmetric bipartition of the system into two parts of length $L=n/2$, any pure state can be parametrized by a matrix $\psi$ as $\ket{\psi} = \sum_{i,j} \psi_{i,j} \ket{i}\otimes\ket{j}$ where indices $i$,$j$ run over configuration states of the subsystems. The Schmidt decomposition reads $\ket{\psi} = \sum_k \sigma_k (\sum_i \alpha_{i,k} \ket{i}) \otimes (\sum_j \beta_{j,k}^* \ket{j})$ where $\psi = \alpha \sigma \beta^H$.
We denote the ground state as $\GS$ and assume it is nondegenerate and thus reflection symmetric
%, $(\GS R)^T = \GS R$, 
and of an even parity. 
%We have introduced a matrix representation $\hat{R} = \sum_{i,j} R_{i,j} \ket{i}\bra{j}$. 
In this notation, it is evident that the reduced density matrix of the system, ${\rm tr}_{\rm right}[\ket{\GS}\bra{\GS}]$, is simply $\GS \GS^H$ and the entanglement entropy, defined as the von Neumann entropy of the reduced density matrix, reads
\(
S^{[\GS]} = - {\rm tr}[ \GS\GS^H \log(\GS\GS^H) ]
\).
In the following we shall be interested in the entanglement of single particle excitations 
$\ket{\Phi} = \hat{X}^\dagger \ket{\GS}$ where $\hat{X}$ is a quasiparticle annihilation operator fulfilling canonical anticommutation relations (CAR)
$\{ \hat{X}, \hat{X}^\dagger \} = \hat{1}$ and $\{ \hat{X}, \hat{X} \} = 0$ and flipping the parity,
$\{ \hat{X}, \hat{P} \} = 0$.
% which requires that operators $\hat{X}\hat{X}^T$ and $\hat{X}^T \hat{X}$ have orthogonal support.
We shall call $\Delta S \equiv S(\Phi) - S(\GS)$ the \emph{excess of entanglement}, as introduced in \cite{berganza}.
Like the ground state, the excited states must also obey the symmetries of the system, being either symmetric or antisymmetric to reflections, $\hat{R} \ket{\Phi} = \pm \ket{\Phi}$, and of an odd parity. Because of these symmetry constraints, there exist maps $\hat{X}_l$ and $\hat{X}_r$, acting only on the left/right part of the system, such that $\ket{\Phi} = \frac{1}{\sqrt{2}}\big( \hat{X}_l^\dagger + \hat{X}_r^\dagger\big) \ket{\GS}$, and $\{\hat{X}_{l,r}, \hat{P} \} = 0$. 
They anticommute $\{ \hat{X}_l , \hat{X}_r \}= \{ \hat{X}_l , \hat{X}_r^\dagger \} = 0$ and thus 
$\{ \hat{X}_{l,r}, \hat{X}_{l,r}^\dagger \} = \frac{1}{2}\hat{1}$ and 
$\{ \hat{X}_l, \hat{X}_l \} = - \{ \hat{X}_r, \hat{X}_r \}$.
% (using CAR for $\hat{X}$).
% which is only possible if $\hat{X}_{l,r}^2 = \pm \epsilon \hat{1}$ for some $\epsilon$.
%

Let us now assume a finite correlation length $\xi$, such that $\vert \langle \hat{x}_i \hat{x}_j^\dagger \rangle  - \langle \hat{x}_i\rangle\langle \hat{x}_j^\dagger\rangle \vert < C e^{-|i-j|/\xi}$ for any two local operators $\hat{x}_{i}$, $\hat{x}_j$ at sites $i, j$. %, in particular for particle annihilation operators. 
Although the quasiparticles are not localized, simultaneous presence in both halves of the systems gets unlikely for $n\to\infty$, 
\begin{equation}
\vert \bra{\GS} \hat{X}_l \hat{X}_r^\dagger \ket{\GS}  \vert < \frac{{\rm const}}{n  \sinh^2(\frac{1}{2} \xi^{-1} ) } = O(n^{-1})
\label{eq:corrineq}
\end{equation}
which results from $\hat{X} = \frac{1}{\sqrt{n}} \sum_{i=1}^{n} (f_j \hat{c}_j + g_j \hat{c}_j^\dagger)$ where $\{ \hat{c}_j \}$ are real particle annihilation operators at sites $\{ j\}$ and $f_j$ and $g_j$ are bounded. The actual quasipartices might be of a different form but they still behave like real particles and this is a valid assumption. 
%In our notation, (\ref{eq:corrineq}) reads ${\rm tr}[ (X_l^H \GS)^H \GS X_r^* ] = O(n^{-1})$.
This bound also results in $\vert\vert X_l \GS\vert\vert_F^2 = \bra{\GS} \hat{X}_l^\dagger \hat{X}_l \ket{\GS} = - \bra{\GS} \hat{X}_l^\dagger \hat{X}_r \ket{\GS} = O(n^{-1})$.

The reduced density matrix $\rho$ for the excited state 
$\Phi = \frac{1}{\sqrt{2}}(\Phi_l + \Phi_r)$ where $\Phi_l = X_l^H\GS$ and $\Phi_r = \GS X_r^*$, now reads
\[
\rho = 
\frac{1}{2}
\Big( 
\Phi_l\Phi_l^H + \Phi_r\Phi_r^H + (\Phi_l\Phi_r^H + \Phi_r\Phi_l^H)
\Big).
\]
The first two terms describe states where either of the two parts is excited and have the same spectrum because of the reflection symmetry. We will show that they are distinguishable whereas the mixed terms vanish in the thermodynamic limit.
Let us first fix the notation, $\vert\vert x \vert\vert_F^2 \equiv \sum_j \sigma_j(x^H x)$, $\vert\vert x \vert\vert_1 \equiv \sum_j \sigma_j(x)$ and $\vert\vert x\vert\vert_\infty \equiv \sigma_1(x)$ where $\sigma_1(x) \geq \sigma_2(x) \geq \dots $ are singular values of $x$.
The distinguishability follows from 
\(
{\rm tr}[  \Phi_l\Phi_l^H\Phi_r\Phi_r^H] \leq 
\vert\vert\GS\vert\vert_F^2 
\vert\vert  X_l \GS X_r^* \vert\vert_F^2
\)
where $\vert\vert \GS\vert\vert_F =1$ and  $\vert\vert X_l \GS X_r^* \vert\vert_F^2 \leq \vert\vert X_r X_r^H\vert\vert_\infty\vert\vert X_l \GS \vert\vert_F^2 = O(n^{-1})$, due to 
$\vert\vert X_l \GS \vert\vert_F^2 = O(n^{-1})$ from~(\ref{eq:corrineq}). 
%due to~(\ref{eq:corrineq}). 
We have used 
${\rm tr}[\mu\nu] \leq \vert\vert \mu \vert\vert_\infty {\rm tr}[\nu] \leq {\rm tr}[\mu]{\rm tr}[\nu]$ for $\mu,\nu \geq 0$ (see e.g. \cite{bhatia}) and  $X_r X_r^H  \leq I \Rightarrow \vert\vert X_r \vert\vert_\infty  \leq 1$. 
The trace norm of the mixed term $\Phi_l\Phi_r^H =X_l^H \GS X_r^T\GS^H = -X_l^H  X_l \GS \GS^H$  also vanishes, 
$\vert\vert X_l^H  X_l \GS \GS^H \vert\vert_1 
\leq \vert\vert (X_l^H  X_l)^2 (\GS \GS^H)^2 \vert\vert_1^{1/2}
\leq \vert\vert \GS\vert\vert_F \vert\vert X_l\vert\vert_\infty \vert\vert X_l\GS\vert\vert_F = O(n^{-1/2})$.
%and $\vert\vert X_l\GS\vert\vert_F = O(n^{-1/2})$ due to~(\ref{eq:corrineq}). 
We have used $\vert\vert \mu\nu\vert\vert_1^t \leq \vert\vert \mu^t \nu^t \vert\vert_1$ for $\mu,\nu > 0$ and $t \geq 1$ \cite{bhatia}. 
It is easy to show that the trace of the mixed norms vanishes as $O(n^{-1})$, again due to~(\ref{eq:corrineq}).
Therefore, %we can write 
\[
\rho = \big(\frac{1}{2} u_l^H \GS\GS^H u_l \big) \oplus \big( \frac{1}{2} u_r^H\GS^H\GS u_r \big) + \rho'
\]
where $\vert\vert \rho' \vert\vert_1 = O(n^{-1/2})$ and $u_{l,r}$ are isometries for which $u_{l,r}^H u_{r,l} =0$. Because $\rho'$ vanishes and ${\rm tr} [u^H \GS\GS^H u] \leq {\rm tr}[\rho]$ for any isometry $u$ and $\rho\geq 0$, whereas on the other hand ${\rm tr}[\rho] = 1$, we conclude
$\sigma_j( u_l^H \GS\GS^H u_l) = \sigma_j( u_r^H \GS^H\GS u_r) = \sigma_j(\GS\GS^H) + O(n^{-1/2})$. Because of the continuity of von Neumann entropy, $\vert S(\rho)-S(\rho')\vert \leq \vert\vert \rho-\rho'\vert\vert_1$, the excess of entanglement for the excited state $\Phi$ approaches $\log(2)$ with corrections of the order of $O(n^{-1/2})$. We will learn from numerical simulations that this bound is not optimal and the corrections are actually of the order of $O(n^{-1})$.
The upper bound $\Delta S \leq \log(2)$ is given by splitting the total Hilbert space $\mathcal{H} = \mathcal{H}_l \otimes \mathcal{H}_r$ and realizing that 
$\ket{\Phi_q}\bra{\Phi_q} = {\rm tr}_{q} \ket{\GS}\bra{\GS}$ since 
 $\ket{\GS}\bra{\GS}$ is separable in the space of quasiparticles $\mathcal{H} = \bigotimes_q \mathcal{H}_q$ ($\ket{\GS}$ being the Fermi sea). Let us now orthogonalize some $\mathcal{H}_q$ with respect to $\mathcal{H}_r$ and write the result as $\mathcal{H}_{q'} \leq \mathcal{H}_l$ such that $\mathcal{H} = \mathcal{H}_{l'}\otimes \mathcal{H}_{q'} \otimes \mathcal{H}_r$. Obviously, $\rho_{l'+q'}$ and $\rho_{l'}$ are reduced density matrices for the ground state and for the quasiparticle excitation, respectively. Using an inequality $S(\rho_{l'}) \leq S(\rho_{l'+q'}) + S(\rho_{q'})$ \cite{nielsenbook} we arrive at $S^{[\Phi_q]}  \leq S^{[\GS]}  + \log(2)$  because $\rho_{q'}$ is of rank 2 and thus $S(\rho_{q'}) \leq \log(2)$.

\textit{Quadratic systems.}
An example of systems where the quasiparticle picture is not only valid but exact, are integrable systems, mappable to free fermions, such as a quantum Ising chain in a transverse magnetic field. We consider a quadratic Hamiltonian in terms of (hermitian) Majorana operators $\{ \hat{w}_i, \hat{w}_j \} = 2 \delta_{i,j}$ on $n$ sites, $\hat{H} = \sum_{i,j=1}^{2n} \hat{w}_i H_{i,j} \hat{w}_j$, where $H=H^H=-H^T$. The $\hat{w}_j$ are chosen such that 
$\big\{ \hat{w}_1^{m_1}\cdots \hat{w}_{2L}^{m_{2L}} ; m_j \in \{0,1\} \big\}$ only act on the Hilbert space of the first $L=n/2$ sites for which they form a complete basis. We diagonalize $H = V \Lambda V^H - V^* \Lambda V^T$ with a $2n \times n$ isometry $V$ for which  $V^H V = I$ and $V^T V=0$,  and define quasiparticle annihilators $\hat{b}_k = \frac{1}{\sqrt{2}} \sum_{i=1}^{2n} V_{i,k}^* \hat{w}_i$, so that we obtain a free fermion model $\hat{H} = \sum_k \varepsilon_k \hat{b}_k^\dagger \hat{b}_k$.
As shown in \cite{latorre,alba}, 
a $2L \times 2L$ correlation matrix
 $\bra{\psi} \hat{w}_i \hat{w}_j \ket{\psi} = \delta_{i,j} + i \Gamma_{i,j}^{[\psi]}$ 
where $\Gamma^{[\psi]}$ is a real skew symmetric matrix with eigenvalues $\pm i  \nu_j^{[\psi]}$, determines the eigenvalues of the reduced density matrix
$\rho^{[\psi]} \equiv {\rm tr}_{\rm right} \ket{\psi}\bra{\psi}$ as
$\rho_{m_1,\ldots,m_L}^{[\psi]} = \prod_{j} \frac{1+m_j \nu_j^{[\psi]}}{2}$ with $m_j=\pm 1$. The entanglement entropy of $\psi$ reads
\(
 S^{[\psi]} = - {\rm tr}\big( \frac{I+i\Gamma^{[\psi]}}{2}  \log \frac{I+i\Gamma^{[\psi]}}{2} \big).
 \) 
Let us denote the upper part of $V$,
referring to sites $\{1,\ldots,L\}$, 
as $u \in \mathbb{C}^{2L\times n}$, 
 where we also reshuffle the columns $\underline{u}_k$  in a way that $\underline{u}_{2k-1}\cdot \underline{u}_{2l} = 0$ and 
$\underline{u}_{2k-1}^* \cdot \underline{u}_{2l-1} = \underline{v}_{2k}^* \cdot \underline{u}_{2l} = \frac{1}{2}\delta_{kl}$, because of the reflection symmetry.
We use $\underline{a} \cdot \underline{b} \equiv \underline{a}^T \underline{b}$.
For the ground state, it is easy to check that $\langle w_i w_j\rangle = 2 [u u^H]_{i,j}$ 
and $\Gamma^{[\GS]} = i (u^* u^T - u u^H)$.
Bases $\{ \sqrt{2} \underline{u}_{2k-1}\}$ and $\{ \sqrt{2} \underline{u}_{2k}^* \}$ are orthonormal
%and thus $\sum_{k=1}^{L} \chi_{2k-1} = \sum_{k=1}^{L} \chi_{2k}$ 
 which allows us to write $\Gamma^{[\GS]}$ as a sum of $L$ rank-2 terms
\[
\Gamma^{[\GS]} = \sum_{k=1}^{L} \chi_{2k-1}
\quad\textrm{for}\quad 
\chi_k \equiv 2 i (
 \underline{v}_{k}^* \otimes \underline{v}_{k}^T - \underline{v}_{k} \otimes \underline{v}_{k}^H   ).
\]
%\( \Gamma^{[\GS]} = \sum_{k=1}^{L} \chi_{2k-1} \) for \( \chi_k \equiv 2 i ( \underline{v}_{k}^* \otimes \underline{v}_{k}^T -\underline{v}_{k} \otimes \underline{v}_{k}^H   ) \).
%
For an excited state $\ket{\Phi_\kappa} = \hat{b}_{\kappa}^\dagger\ket{\GS}$, it is easy to see that 
$\Gamma^{[\Phi_\kappa]} = \Gamma^{[\GS]} - \chi_{\kappa} = \sum_{k \neq \kappa'} \chi_{2k-1}$,  
assuming $\kappa = 2 \kappa'-1$,
which is obviously rank deficient with $\textrm{dim}\,\textrm{ker}\Gamma^{[\GS]} = 2$. This makes the spectrum of the reduced density matrix double degenerate. The same reasoning is applied to
$\hat{b}_{2\kappa'-1}^\dagger \hat{b}_{2\mu'-1}^\dagger \ket{\GS}$ (or $\hat{b}_{2\kappa'}^\dagger \hat{b}_{2\mu'}^\dagger \ket{\GS}$), resulting in a four times degenerate spectrum of the reduced density operator.
Let us now show that  $\Delta S$ approaches $\log(2)$.
Using the Taylor expansion, we write
\[
\Delta S = \sum_{j=1}^{\infty} [2j (2j-1)]^{-1} 
{\rm tr}\big[
(i\Gamma^{[\GS]})^{2j} - (i\Gamma^{[\GS]} - i\chi_\kappa)^{2j} \big] 2^{-1}
\]
where ${\rm tr}[(i\chi_k)^{2j}] \leq 2$. If 
$(\forall k,l)\, \underline{u}_k\cdot\underline{u}_l = 0$, then the trace in above expression equals $2$, since $\chi_k \chi_l = \delta_{kl} \chi_{k}^2$,  and thus $\Delta S = \log(2)$.
Using $\hat{b}_{k; L} = \frac{1}{\sqrt{2}} \sum_{i=1}^{2L} u_{i,k}^* \hat{w}_i$ we see that this is indeed the case for $n\to\infty$ where $\sum_l \vert \underline{u}_k\cdot \underline{u}_l \vert^2 = \bra{\GS} \hat{b}_{k; L}^\dagger \hat{b}_{k; L} \ket{\GS} = O(n^{-1})$ due to~(\ref{eq:corrineq}). Hence, the excess of entanglement for a single quasiparticle excitation equals $\Delta S = \log(2)$ and for $k$-quasiparticle excitation it equals $k \log(2)$ (if all excited quasiparticles belong to the same reflection symmetry class).

\textit{Results.}
%%%%%%%%%%%%%%%%%%%%%%%%%%%%%%
% TILTED ISING MODEL  
%%%%%%%%%%%%%%%%%%%%%%%%%%%%%%
From a very generic setting we have seen that the entanglement entropy of excited states is connected with the number of quasiparticles which we shall now confirm by numerical simulations. First we consider an antiferromagnetic Ising chain in a mixed transverse and longitudinal magnetic field, the so-called tilted Ising chain, described by a Hamiltonian operator $H = \sum_{j} \big(\sigma_j^x \sigma_{j+1}^x + \sigma_j^z + \sigma_j^x\big)$ where we have chosen a non-critical (gapped) regime featuring strong non-integrable effects. The model is not exactly solvable and we rely on the density matrix renormalization group \cite{white} to calculate the first few lowest energy excited states. To check the validity of the results, we increase the bond dimension and optimize the results by a variational optimization technique \cite{varnrg}. Eventually we settle on $D=100$. The states are rotated in a way that they are eigenstates of the reflection operator. We calculate the entanglement entropies for $M=50$ lowest energy excited states and various chain lengths, $n=50,100,200,500$.
\begin{figure}
\centering
\includegraphics[width=\columnwidth]{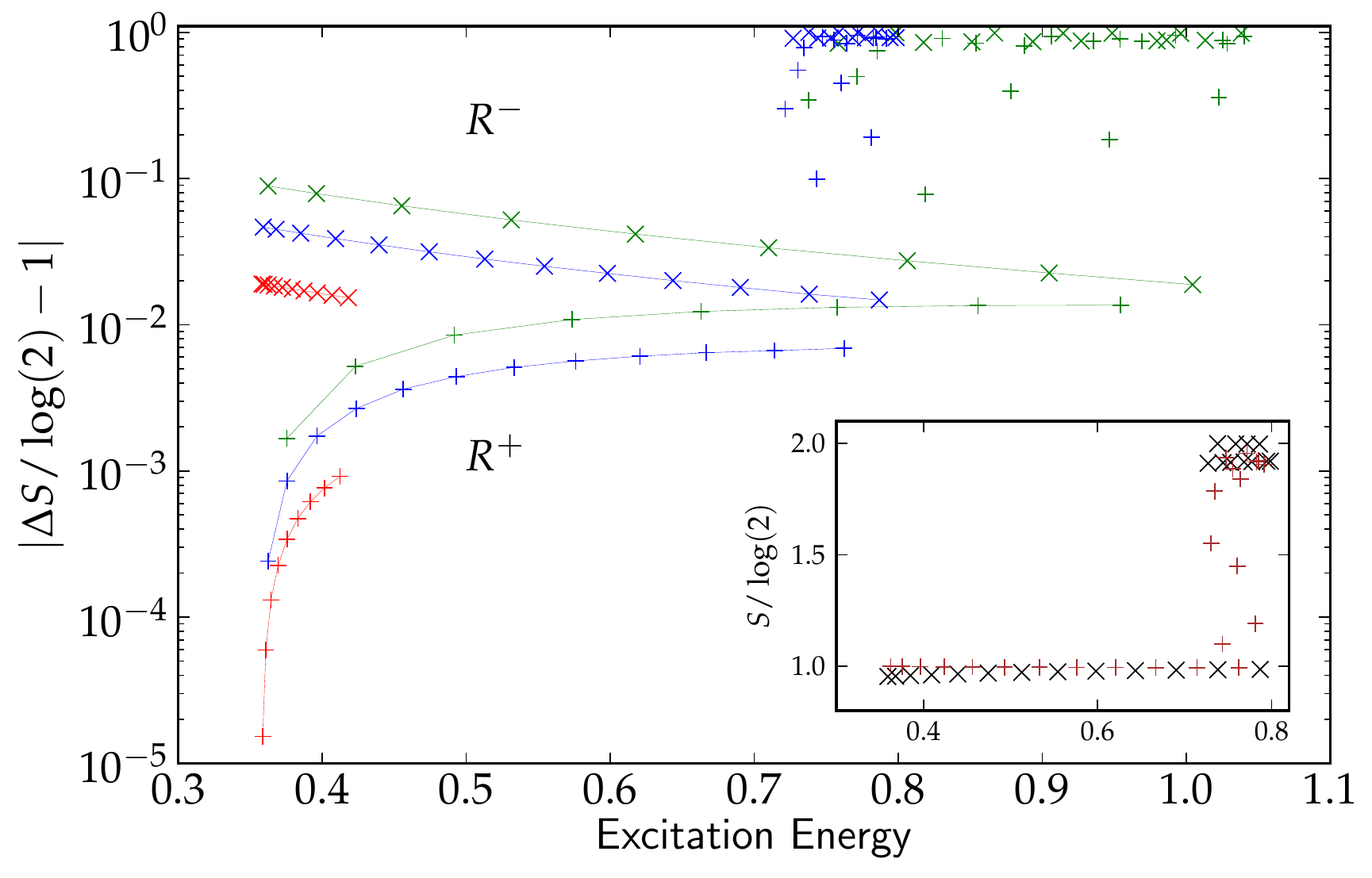}
\caption{Excess of entanglement $\Delta S$ for the tilted Ising chain on $n=100,200,500$ (top to bottom; green, blue, red) sites for reflection symmetric (plus) and antisymmetric (cross) states.
We have connected points where $\vert \Delta S/\log(2) -1 \vert < 0.1$.
The inset shows $\Delta S$ vs energy for $n=200$.
}
\label{fig:tiltedising}
\end{figure}
From Fig.~\ref{fig:tiltedising} we observe that the excess of entanglement approaches $\Delta S \to \log(2)$ for the lowest exited states which is consistent with single quasiparticle excitations; reflection symmetric and antisymmetric excitations display distinct behavior.
We observe another clustering at $\Delta S \sim 2\log(2)$ which corresponds to two-quasiparticle excitations. Note that for $n=500$, all states that we calculate using the DMRG, are single particle excitations since the two-particle excitations only start at energy approximately $0.7$ (i.e. two lowest single particle excitations).
Above this threshold, we also observe reflection invariant states with irregular behavior which, we conclude, must be bound states.
%It is interesting to note that low-temperature physics is largely unaffected by the nonintegrability. 

%%%%%%%%%%%%%%%%%%%%%%%%%%%%%%
% XY MODEL 
%%%%%%%%%%%%%%%%%%%%%%%%%%%%%%
We do a systematic study of excitations for an exactly solvable Heisenberg XY model, described by a Hamiltonian
$H = \sum_{j} \frac{1+\gamma}{2} \sigma_j^x \sigma_{j+1}^x + \frac{1-\gamma}{2}\sigma_j^y\sigma_{j+1}^y + \sum_j h \sigma_j^z$ where we choose a particular point in the parameter space $(\gamma,h) = (0.5,0.9)$;  any point with $h\neq 1$ and $\gamma \neq 0$ gives similar results.
\begin{figure}
\centering
\includegraphics[width=\columnwidth]{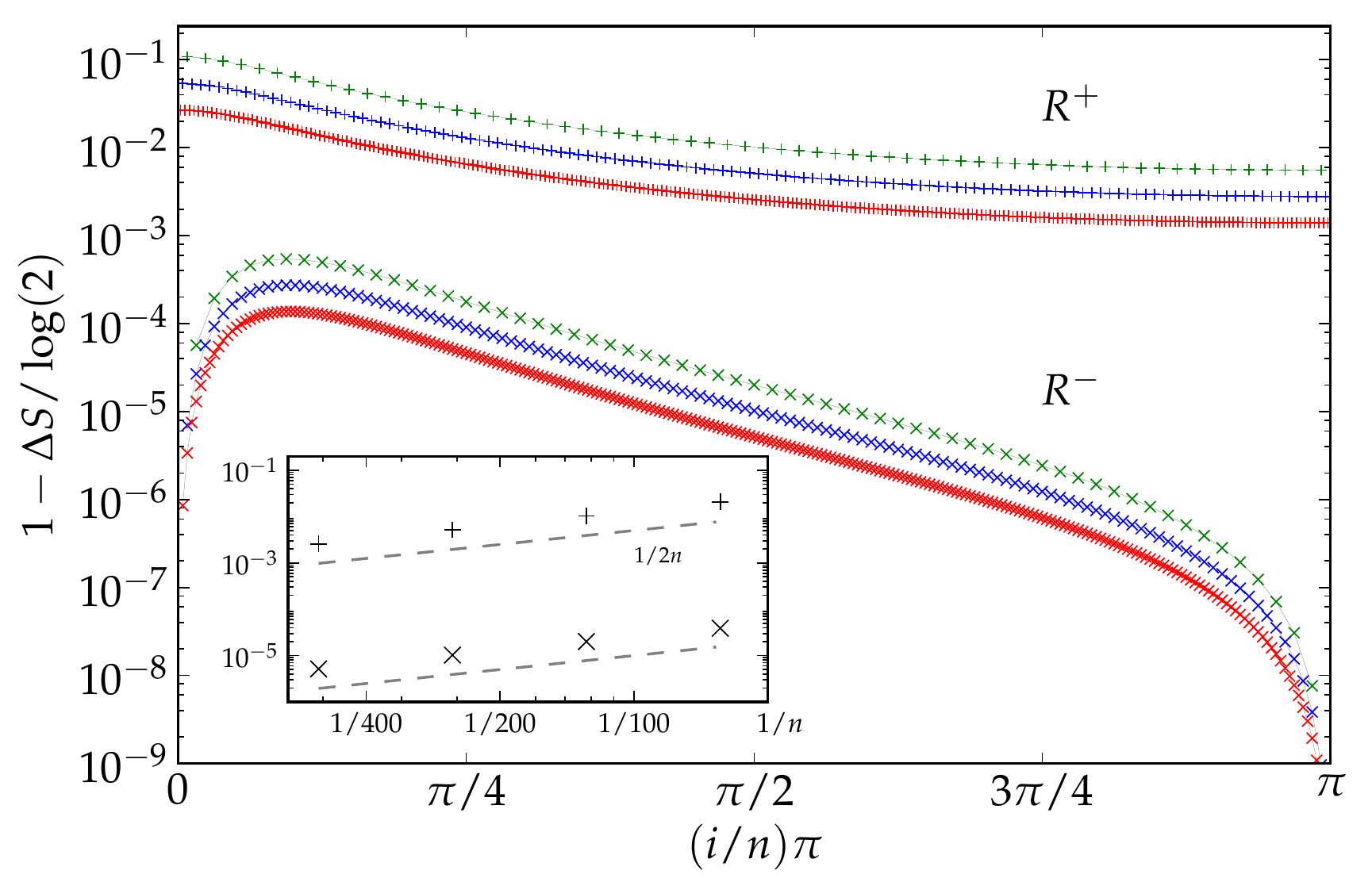}
\caption{Excess of entanglement $\Delta S$ of single quasiparticle excitations $\hat{b}_i^\dagger \ket{\GS}$ of the XY($\gamma=0.5,h=0.9$) model for $n=128,256,512$ (top to bottom; green, blue, red) and reflection symmetric ($+$, plus) and antisymmetric ($-$, cross) states.
The inset shows the scaling of $\Delta S$ with $n$ at phase $\pi/2$ for $R^\pm$ (plus/cross).}
\label{fig:xy}
\end{figure}
In Fig.~\ref{fig:xy} we show the excess of entanglement of all single particle excitations $\hat{b}_{i}^\dagger \ket{\GS}$ for chain lengths $n=128,256,512$ and again observe that it approaches $\Delta S \to \log(2)$. The scaling analysis confirms that $\log(2) - \Delta S$ scales as $n^{-1}$ as shown in the inset.

%%%%%%%%%%%%%%%%%%%%%%%%%%%%%%
% XY MODEL  : THREE PARTICLE EXCITATIONS
%%%%%%%%%%%%%%%%%%%%%%%%%%%%%%
Three-particle excitations are studied on the case of an Ising chain in a transverse field (i.e. the XY model with $(\gamma,h) = (1,2)$). We choose excitations of a form $\ket{\Phi} = \hat{b}_{i}^\dagger \hat{b}_{k_1}^\dagger \hat{b}_{k_2}^\dagger \ket{\GS}$ where $k_1$ and $k_2$ are fixed and correspond to reflection antisymmetric excitations at quasi-momenta $\pi/4$ and $3 \pi/4$. 
\begin{figure}
\centering
\includegraphics[width=\columnwidth]{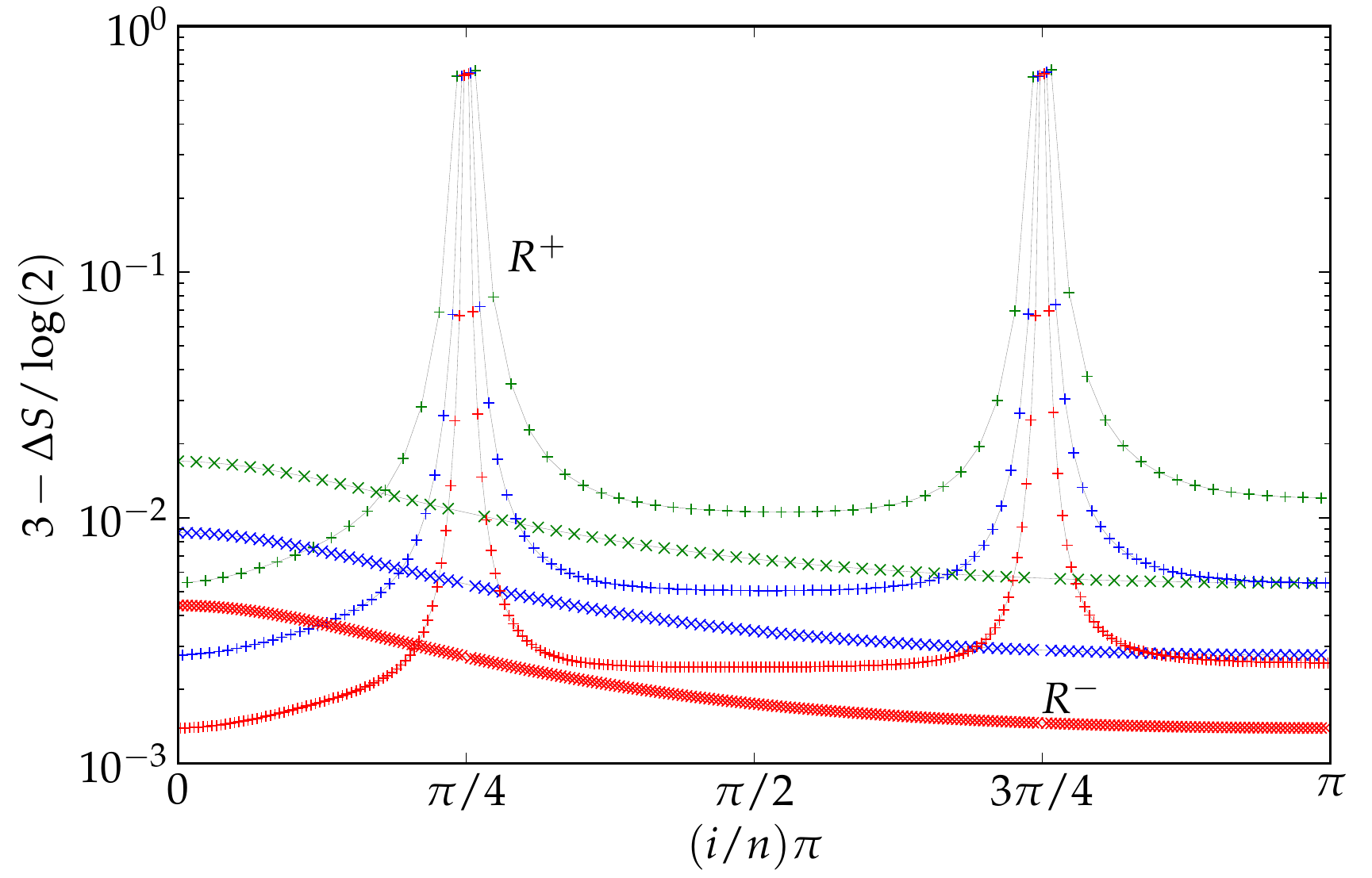}
\caption{Excess of entanglement $\Delta S$ of $\hat{b}_{i}^\dagger \hat{b}_{ 3n/4}^\dagger \hat{b}_{n/4}^\dagger\ket{\GS}$ for the Ising model ($h=2$) with $n=128,256,512$ (top to bottom; green, blue, red) for reflection symmetric ($R^+$, plus) and antisymmetric ($R^-$, cross) states.}
\label{fig:isingthree}
\end{figure}
We observe in Fig.~\ref{fig:isingthree} that $\Delta S \to 3 \log(2)$ for $n\to\infty$ except for two peaks at $i \sim k_1$ and $i \sim k_2$ where two quasiparticles can form a bound state and the excited state looks like a two-particle excitation with the excess of roughly, but more than $2 \log(2)$, as the numerical results suggest.

\textit{Discussion.}
The main message of this Letter, that the entanglement entropy of single particle excitations exceeds the the ground state for $\log(2)$, can be used to justify the Ansatz for single particle excitations \cite{jutho}. They are based on a homogeneous matrix product state for the ground state described by a tensor $A^{s}\in \mathbb{C}^{D\times D}$ as $\ket{\GS} = \sum_{\underline{s}} v_l ( \cdots A^{s_{-1}} A^{s_0} A^{s_1} \cdots )v_r \ket{\underline{s}} $, see \cite{tdvp}, and are given as a \emph{sum} of matrix product states where all tensors are equal to $A^{s}$ except for a site $j$ where some other tensor $B^{s}$ is used, together with a phase factor $e^{\kappa j}$ for momentum $\kappa$. Formally, the Ansatz reads
\(
\ket{\Phi_\kappa(B)} = \vert \mathbb{Z} \vert^{-1/2} \sum_{m \in \mathbb{Z}; \underline{s} } 
e^{i \kappa m} \hat{T}^{m} v_l^\dagger ( \cdots A^{s_{-1}} B^{s_0} A^{s_1} \cdots )v_r \ket{\underline{s}}
\).
For each $\kappa$, there exist $D^2 (d-1)$ ($d=2$ for spin-$1/2$) orthonormal states $\ket{\Phi_\kappa(B)}$, such that $\braket{\GS}{\Phi_\kappa(B)}=0$. The relevant quantities in infinite matrix product states are the left and the right eigenvector of a transfer matrix $E = \sum_s A^s \otimes {\overline A}^{s}$, corresponding to the largest eigenvalue, which can be reshaped to hermitian positive definite $D\times D$ matrices $l,r$. An overlap of semi-infinite states with respect to a bipartite splitting, $\GS = \alpha \beta^T$ in the notation of the first part of the Letter, is given exactly by these matrices as $\alpha^H \alpha = l$ and $\beta^H \beta = r^T$, and so is the reduced density matrix which reads $\rho = \alpha \beta^T\beta^* \alpha^H = {\tilde\alpha} \Xi {\tilde\alpha}^H$ where $\Xi = L^H r L$ and ${\tilde\alpha} = \alpha L^{-H}$ such that ${\tilde \alpha}^H{\tilde\alpha}= I$. We have used a Cholesky decomposition $l = L L^H$.
The excitations can be written as  $\Phi = \frac{1}{\sqrt{2}}( \phi \beta^T + \alpha \theta^T)$ where the first/second part contains all excitations on the left/right part of the chain. It is easy to show that also $\phi^H\phi = l$ and 
$\theta^H \theta = r^T$ whereas $\theta^H \alpha = \phi^H\beta = 0$ (and $\phi^H\alpha = 0$ by construction) which means that $\phi$ is orthogonalized by the same transformation as $\alpha$, that is ${\phi} = {\tilde\phi} L^{-H}$. The reduced density operator reads $\rho = \frac{1}{2}(\phi r \phi^H + \alpha r\alpha^H)$ and 
we immediately obtain 
\(
\rho = \frac{1}{2}( {\tilde\phi} \Xi {\tilde\phi}^H + {\tilde\alpha}\Xi{\tilde\alpha}^H )
\) where ${\tilde\phi}$ and ${\tilde\alpha}$ are isometries with orthogonal support, ${\tilde\alpha}^H {\tilde\phi}=0$, and thus $\rho \sim \frac{1}{2} (\Xi \oplus \Xi)$. 
The entanglement entropy of the Ansatz is equal to $S^{[\Phi]} = S^{[\GS]]} + \log(2)$, irrespectively of the bond dimension $D$, which we have shown to agree with single quasiparticle excitations in the thermodynamic limit. By this we provide additional support that the Ansatz in \cite{jutho} is a good Ansatz for describing single particle excitations. Unfortunately, however, the entanglement entropy of almost all other excited states (except e.g.  some bound states) goes beyond this limit; in this case a generalized Ansatz should be used where not one but two or more tensors in $A^{s}$ the ground state are replaced by some $B^{s}$. 

%We also note that in case the ground state is separable, such as for the XX model in magnetic field, the entanglement entropy of all single particle excitations is $\log(2)$ for any system size.

\textit{Conclusion.}
We have studied the entanglement entropy of single quasiparticle excitations and shown that it is related to the ground state entanglement entropy with a contribution of $\log(2)$ for each quasiparticle in the thermodynamic limit, assuming a finite correlation length and a valid quasiparticle description at low energies. We have provided numerical evidence for this universal behavior which suggests a possibility to classify excited states on the basis of the entanglement.

\textit{Acknowledgments.}
Discussions with F. Verstraete and T. Prosen are gratefully acknowledged, as well as financial support by the EU project QUEVADIS. The computational results were in part achieved using Vienna Scientific Cluster.

\end{document}